\documentclass[9pt]{article}
\usepackage{spconf}
\usepackage{blindtext}
\usepackage{cite}
\usepackage{graphicx}
\usepackage{amsfonts}
\usepackage{color}
\usepackage{xcolor}
\usepackage{amsmath}
\usepackage{stmaryrd}
\usepackage{algorithm, caption}
\usepackage{algpseudocode}
\usepackage{hyperref}
\usepackage{amsthm} 
\usepackage{mathtools}
\usepackage{subcaption} 
\usepackage{bm}
\usepackage{booktabs}
\usepackage{colortbl} 
\usepackage{booktabs}
\usepackage{float}
\usepackage{siunitx}
\usepackage{fixltx2e}
\usepackage{multirow}
\usepackage[T1]{fontenc}
\graphicspath{{Figures/}}

\DeclarePairedDelimiter\abs{\lvert}{\rvert}%
\DeclarePairedDelimiter\norm{\lVert}{\rVert}%

\makeatletter
\let\oldabs\abs
\def\abs{\@ifstar{\oldabs}{\oldabs*}}
\let\oldnorm\norm
\def\norm{\@ifstar{\oldnorm}{\oldnorm*}}
\makeatother

\hypersetup{
	bookmarksnumbered=true, bookmarksopen=true, colorlinks,
	citecolor=black,%
	filecolor=black,%
	linkcolor=black,%
	urlcolor=black
}

\begin{document}

\title{Hearables: Ear EEG based Driver Fatigue Detection}
\name{Metin C. Yarici, Pierluigi Amadori, Harry Davies, Takashi Nakamura, Nico Lingg, \\ Yiannis Demiris, Danilo P. Mandic}
\address{Department of Electrical and Electronic Engineering, Imperial College London, SW7 2AZ, UK
\\E-mails: \{metin.yarici16,  d.mandic\}@imperial.ac.uk}

\markboth{Journal of \LaTeX\ Class Files,~Vol.~6, No.~1, January~2007}%
{Shell \MakeLowercase{\textit{et al.}}: Bare Demo of IEEEtran.cls for Journals}

\maketitle

\begin{abstract}
Ear EEG based driver fatigue monitoring systems have the potential to provide a seamless, efficient, and feasibly deployable alternative to existing scalp EEG based systems, which are often cumbersome and impractical. However, the feasibility of detecting the relevant delta, theta, alpha, and beta band EEG activity through the ear EEG is yet to be investigated. Through measurements of scalp and ear EEG on ten subjects during a simulated, monotonous driving experiment, this study provides statistical analysis of characteristic ear EEG changes that are associated with the transition from alert to mentally fatigued states, and subsequent testing of a machine learning based automatic fatigue detection model. Novel numerical evidence is provided to support the feasibility of detection of mental fatigue with ear EEG that is in agreement with widely reported scalp EEG findings. This study paves the way for the development of ultra-wearable and readily deployable hearables based driver fatigue monitoring systems. 
\end{abstract}



\section*{Introduction}

Road accidents can lead to serious injuries for passengers and economic losses in the transport system. According to the Department for Transport in the UK, the total cost of reported accidents in the UK was approximately £36 billion in 2016. Among car accidents, 43\%  (59,586 out of 137,798) of accidents were reported due to a "Driver error or reaction", such as when a driver does not perceive potential hazards or miscalculates the speed or trajectory of other cars \cite{transport_tech_rep_2016}. These human errors are often associated with driver fatigue, which can cause a lapse in vigilance and the onset of sleep \cite{bioulac2017risk}. A recent study found that the prevalence of falling asleep at the wheel was 17\% in Europe, with 13\% of consequent accidents which were caused as a result of the fatigued driver leading to hospitalisation \cite{aakerstedt2013white}. 

Two main categorises of driver fatigue exist: i) sleep related fatigue and ii) task related fatigue. Sleep-related fatigue is caused by a lack of sleep or poor quality sleep, whereas task related fatigue is caused by driving itself and by the driver's environment. Experts further differentiate between active and passive task related fatigue. Active task related fatigue occurs as a result of an accumulation of cognitively demanding tasks, such as navigating through poor visibility \cite{aakerstedt2013white}, and is the most common type of driver fatigue. On the other hand, passive task-related fatigue arises due to prolonged, monotonous exertion, for example during long haul driving in low traffic environments. Passive task related fatigue causes more road accidents than active task related fatigue, despite a lower prevalence\cite{saxby2008effect}. Typical driving impairments that are experienced as a result of driver fatigue are slowed reaction time, drifting between lanes, and difficulty gauging the speed of other vehicles \cite{dinges1997cumulative,horne1995sleep,lal2001critical}. Additionally, driver fatigue can lead to an inability to accurately assess one's own fitness to continue driving \cite{brown1997prospects}.

Driver fatigue monitoring is therefore a pre-requisite to detect and quantify signs of fatigue in a driver before it becomes a serious safety risk \cite{johns2008new}. Common techniques used for this purpose include measuring the driver's behaviour, such as speed regulation, break reaction time, and steering action \cite{langner2010mental,zhang2016sensitivity}. Additionally, video cameras and other sensors can be used to identify increased frequency of blinking, yawning, and head nodding \cite{dreissig2020driver,fan2007yawning}. Other methods include subjective self-assessments (questionnaires) \cite{lal2001critical}, and physiological measurements such as heart rate \cite{lu2022detecting}, EEG (Electroencephalography) or other brain activity measurements \cite{lal2002driver, nguyen2017utilization}. Behavioural monitoring fails to provide a direct measure of the drivers mental and physical state, while subjective self-assessment is difficult to implement in real-time, and is susceptible to a fatigued driver's inability to assess their own fitness. Physiological monitoring is a popular method for driver fatigue assessment as a result of its ability to provide a direct and continuous measure of the mental and physical state of the driver.

Of the multiple physiological monitoring methods available, brain monitoring via EEG is a popular choice for driver fatigue monitoring as it can provide a direct measure of the driver’s mental state \cite{lal2003development,subha2010eeg,monteiro2019using}, which can be used to make more informed decisions about fatigue management. In addition, EEG is a non-invasive and relatively easy way to measure brain activity, and has been shown to be an effective method for mental fatigue detection \cite{langner2010mental, lal2002driver, tran2020influence,caldwell2002eeg, cao2014objective, fan2015electroencephalogram, jagannath2014assessment, jap2009using, tanaka2012effect, trejo2015eeg, golz2007feature, sommer2005fusion}.

\begin{figure}[!h]
    \centering
    \includegraphics[width = \columnwidth]{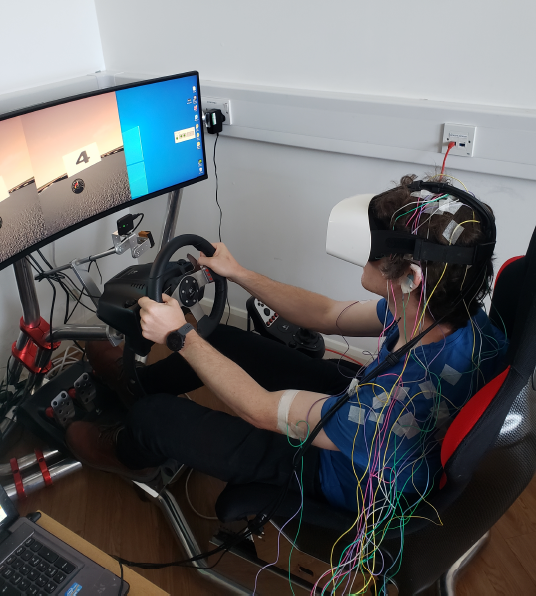}
    \caption{Experimental setup. A subject seated in the physical simulator during the driving task. The screen displays a live stream of the view of the simulated environment from the perspective of the driver (as viewed from within the VR headset). The physical simulator comprises a steering wheel, seat, and accelerator pedal.}
    \label{fig: setup}
\end{figure}

However, driver fatigue monitoring via conventional EEG equipment is not feasible as a result of the cumbersome nature of the equipment; this requires the placement of sensors on the scalp. Additionally, conventional EEG hardware can be obtrusive, time consuming to set-up, and difficult to operate without specialist knowledge. In contrast, wearable EEG technology is designed with the comfort of the user and ease of use in mind, and is therefore well suited to driver fatigue monitoring \cite{casson2010wearable}. Unlike conventional EEG, wearable EEG usually comes in the form of a lightweight head band which is comfortable to wear over long periods of time. Several studies have demonstrated the feasibility of wearable EEG-based driver fatigue monitoring \cite{li2015context, lin2014wireless, zhu2021vehicle}. Recently, wearable EEG devices that record from the position of the ear (ear EEG), the so-called 'hearables', have also been introduced, whereby ear-worn devices, such as in-ear phones, or hearing aids, serve as a miniature, lightweight base for wearable EEG technology \cite{mikkelsen2015eeg, looney2012ear}. Advantageously, hearables (and ear EEGs) are lightweight, comfortable to wear, portable, minimally impeding, and easy to use, making them well suited to driver fatigue monitoring. Moreover, in addition to brain monitoring, the detection of multiple other forms of vitals sign from the position of the ear have recently been established, for example the detection of heart rate, breathing rate, and blood oxygenation \cite{davies2020ear,davies2022wearable,goverdovsky2017hearables} and ECG \cite{von2017hearables}. Furthermore, the ability to provide environmental context, through classification of activities (such as talking, walking, and eating) during multi-modal ear-EEG recordings has also been demonstrated \cite{goverdovsky2014co,hammourArt,OcchipintiArt}; reliable separation of EEG from measurement noise, in addition to identification of the users activity, can greatly increase the utility and robustness of a wearable EEG fatigue monitoring system. Overall, 'hearable' devices hold much promise for the purpose of driver fatigue monitoring.

However, limited research on on ear-EEG based driver fatigue monitoring exists. Hwang \textit{et al.} investigated the utility of a hybrid ear-scalp EEG system, whereby an in-ear EEG electrode was referenced to a conventional scalp-EEG electrode, and found that classification of fatigued and alert periods exceeded that of single-channel conventional scalp EEG. The authors also showed that spectral power in the theta range (\SI{4}{Hz} - \SI{8}{Hz}) increased in both the hybrid ear-scalp EEG and conventional scalp EEG. However, the authors did not employ artifact rejection methods, which are required for rigorous EEG assessment, while the utility of a scalp electrode in conjunction with an ear electrode reduces the wearability of the fatigue monitoring system. Other studies have investigated the use of solely ear-based ear EEG in the context of driver fatigue related phenomena, such as sleep onset detection \cite{nakamura2018automatic} and sleep stage tracking \cite{stochholm2016automatic, mikkelsen2019accurate, nakamura2019hearables}. The aim of this paper is to establish the feasibility of detecting passive, task-related mental fatigue in a driving scenario using a solely ear-EEG based set up. Measurements on ten subjects in a monotonous simulated driving task were conducted. Widely reported changes in scalp EEG were reflected well in the ear EEG measurements, while automatic classification of alert and fatigued phases via a machine learning model demonstrated the feasibility of real-time, ear EEG based driver fatigue monitoring. 

In agreement with widely reported findings, false discovery rate (FDR) adjusted increases in frontal theta band (Cohen's \textit{d} = 1.3, 95\% confidence interval (CI) = [0.1, 0.9], p-value (\textit{p}) = 0.046) and posterior alpha band EEG (\textit{d} = 3.2, CI = [1, 1.5], \textit{p} < 0.001) are reproduced in this study. Between alert and fatigued states, the proposed ear EEG system also showed statistically significant increases in theta band EEG (\textit{d} = 1.3, CI = [0.2, 0.9], \textit{p} = 0.0013), but not in the alpha band (\textit{d} = 0.8, CI = [0.1;0.8], \textit{p} = 0.066). Binary classification of \SI{10}{s} long EEG segments from alert and mental fatigue phases based on a single ear EEG channel, with validation accuracy, (acc. = 70\% ) and a Matthews Correlation Coefficient (MCC = 0.4), performed on par with a multi-channel (Fz, Cz, and POz) scalp EEG model (acc. = 74\%, MCC = 0.48). The findings in this paper open new avenues in seamless, reliable, and efficient, hearable EEG based driver fatigue monitoring.

\section*{Methods}
Ten participants (mean age 24, 8 male) were recruited for this study, and provided written informed consent before the presented experimentation was conducted. All experimentation were conducted in accordance with the Declaration of Helsinki, under the Imperial College London ethics committee approval JRCO 20IC641. 

\subsection{Simulated monotonous driving}
Previous research suggests that monotonous driving is a main contributor to mental fatigue experienced by drivers \cite{aakerstedt2013white, korber2015vigilance,brown1997prospects}. Key features of monotonous driving are a low level of cognitive demands, a repetitive nature, and a prolonged duration. All of these features were included in a simulated monotonous driving task that each participant was required to complete while measurements of EEG were conducted. The simulation was constructed in the Unreal Engine (https://www.unrealengine.com/), and comprised a circular track with a large enough radius to ensure a low difficulty in driving (no sharp turning), with minimal distraction (no traffic or road signs), and a simple task to occupy the drivers over the prolonged duration. Drivers were instructed to maintain a central position in the road and collide with visual/non physical barriers placed periodically in the centre of the road around the track. The drivers were also instructed to maintain a constant speed at all times; importantly the speed was moderate and the driving task was not difficult. Participants were allowed a brief warm-up period prior to measurements, during which a familiarity with the simulator could be acquired. The main trial lasted one hour, over which the steering wheel data, and EEG data were continuously monitored. Since previous studies have suggested that mental fatigue can manifest as soon as five minutes after the commencement of a monotonous task, the EEG data acquired from the first five minutes of the trial was treated as 'alert EEG', while the final fifteen minutes were treated as 'mentally fatigued EEG'. Steering wheel behaviour across participants indicates that these periods were indeed reflective of the alert and mentally fatigued states of the drivers (Figure \ref{fig:psych. bahv.}) \cite{zhang2016sensitivity}. In accordance with guidelines for monitoring mental fatigue in controlled conditions, participants were required to abstain from alcohol and caffeine consumption, respectively, for twenty-four and twelve hours prior to the commencement of the experiment. Each experiment was conducted between the times of 11AM and 5PM (including experimental setup of physiological signal acquisition).

\subsection{Physiological signal acquisiton}
In order to validate the ear EEG, benchmark scalp EEG was simultaneously recorded. In this way, characteristic signs of mental fatigue could be identified in both the proposed ear EEG and conventional scalp EEG system. All EEG measurements were acquired through a g.tec gUSBamp (2011) bio-amplifier, at a sampling rate of \SI{1.2}{kHz}. The GRASS Ag/AgCl reusable cup electordes were used to measure the scalp EEG and alos provide a reference a nd ground for the ear EEG. Ag/AgCl electrodes were mounted at the frontal (Fz), central (Cz), posterior (POz), and temporal (T7,T8) scalp positions in accordance with the standard 10-20 convention. A reference electrode was positioned on the left helix, and a ground electrode was positioned on the left ear lobe. An additional electrode was positioned on the right helix for the purpose of obtaining right ear EEG. In this way, the referencing system did not bias the ear EEG with scalp EEG signals. Ear EEG must be comfortable to wear over prolonged periods, therefore, an alternative to the rigid Ag/AgCl cup electrodes is required for outer ear measurements, as a result of the highly curved and small skin surface of the outer ear. Therefore, custom flexible electrodes, mounted on malleable silicone, were used for this study. The flexible electrodes are made out of a conductive textile substrate (silver coated nylon interwoven with elastic fibres), and have previously been shown to provide stable EEG measurements (low impedance) over the course of a normal working day for five subjects \cite{goverdovsky2017hearables}. A custom flexible electrode was positioned on the concha of the left and right ears of each participant. In this way, three ear EEG configurations were measured; left ear (left concha referenced to left helix), right ear (right concha referenced to right helix), and cross-ear (right concha referenced to left helix). In addition to EEG, a reference ECG signal was recorded by placing an Ag/AgCl electrode on the left arm (referenced to the left helix). A vertical electro-oculogram (VEOG) signal was also recorded from above and below the left and right eyes (referenced to the left helix). 

\subsection{EEG processing}
All signal processing was conducted in MatLab. All channels were first bandpass-filtered between \SI{1}{Hz} and \SI{30}{Hz}, using the MatLab function \texttt{butter} (order 3), before a visual inspection for artifacts. Channels which displayed excessive corruption were removed from the analysis. For each participant, the temporal scalp EEG (T7, T8) and the single ear EEG (left ear and right ear) were excessively corrupted by noise from motion of the user, muscle artifacts (EMG), and the motion of the VR headset (specifically the T7 and T8 electrodes). Consequently, these channels were excluded from the analysis, resulting in one ear EEG channel available for the analysis (cross-ear EEG) and the standard frontal (Fz), central (Cz), and posterior (Pz), scalp EEG channels. Although the single ear EEG and cross-ear EEG channels comprised the same electrodes, the large inter-electrode distance (and increased EEG amplitude) resulted in a sufficient SNR for the cross-ear channel in the challenging real-world measurement scenario. While single ear EEG is most desirable (as a result of the requirement of only a single device), the cross-ear EEG configuration (comprising an earpiece in each ear) is still highly wearable and suitable for driver fatigue monitoring. 

\begin{figure*}[!h]
    \centering
    \includegraphics{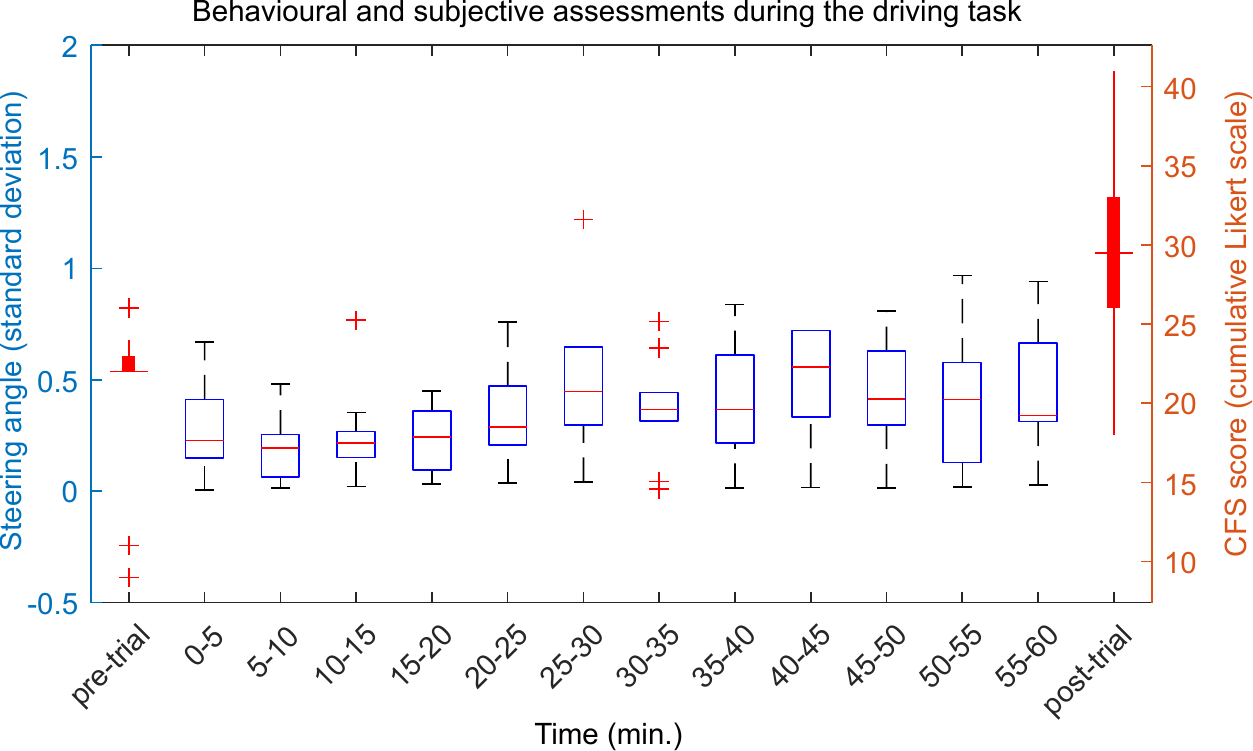}
    \caption{Steering wheel manoeuvres (angle) during driving and pre- and post-trial Chalder Fatigue Scale (CFS) fatigue assessment results. (Left y-axis / unfilled boxes) Distribution of the standard deviation of the steering wheel angle across subjects for five minute periods in the driving trial. (Right y-axis / filled red boxes) Distribution of CFS scores across subjects at pre-trial and post-trial instances. CFS scores of 22-34 indicate mild-to-moderate fatigue, while scores above 35 indicate severe fatigue.}
    \label{fig:psych. bahv.}
\end{figure*}

Following the manual rejection of channels, the remaining channels were conditioned in three more stages: 1) removal of blink artifacts, 2) removal of ECG artifacts, and 3) amplitude threshold rejection. In stage 1), blink artifacts were removed from the EEG by calculating a channel-specific artifact template, and removing this sequentially at each point in the recording where a blink occured (whereby blinks were identified in the VEOG channel). Details of the procedure are provided in the supplemental material, in addition to spectral coherence analysis between the EEG and VEOG channel, demonstrating the efficacy of the artifact removal method. Stage 2) employed a similar procedure, whereby a channel-specific cardiac cycle (ECG waveform) was subtracted from the EEG at each instance of a cardiac cycle, as identified in the reference ECG channel. The details of this procedure, and artifact removal evaluation, are provided in the supplemental material. Stage 3) consisted of a simple artifact rejection procedure, whereby th EEG data was split into 1 second long epochs, and assessed for excessive amplitude (artifacts). The amplitude threshold was set to \SI{200}{\mu V} for the cross-ear EEG and the Fz, Cz, and POz, channels. For all subjects, this resulted in a rejection rate below \SI{10}{\%}.

\subsection{Subjective assessment of mental fatigue}
In addition to physiological measures of mental fatigue, the Chalder Fatigue Scale (CFS), a self-rating scale used to assess the severity of fatigue in adults, was used to assess the level of mental fatigue  \cite{chalder1993development}. The CFS includes a series of eleven questions regarding subjective feelings of fatigue which the participant provides a score for (1-4, low to severe). The total score from all eleven questions is used to determine the level of fatigue in the participant. The CFS evaluation was conducted prior to and after the one hour, monotonous driving task in order to evaluate the level of mental fatigue in the participants that was induced by the driving task. Results for the CFS scoring are displayed in Figure \ref{fig:psych. bahv.}.

\section*{Results}

The analysis of steering wheel behaviour of drivers can help identify signs of fatigue \cite{zhang2016sensitivity}. While experiencing fatigue, a driver's attention to the road can decrease, resulting in the car wandering and drifting between lanes. Such driving is reflected in large angle steering manoeuvres. In the current study, drivers were instructed to maintain a central position in the road throughout the trial, and to pass through visual check-points in the centre of the road. Additionally, this was all achieved while maintaining a constant speed. As such, the requirements for mental fatigue (monotony, low level cognitive demands, prolonged duration) were achieved during the task, and the steering wheel manoeuvres of the driver during the task could therefore provide a behavioural measure of mental fatigue. 

Figure \ref{fig:psych. bahv.} displays the standard deviation of the angle of the steering wheel (rotation) in five minute periods starting from the beginning of the trial up to the end. The box-plots show the distribution of data from all subjects. The steering wheel angles from each subject were normalised (z-score) for the entire trial; in this way, the inter-individual differences in steering style could be accounted for. As expected, there was a general increase in the standard deviation of the steering wheel angle towards the end of the trial. A small dip between the first and second five minute period can be explained by a period of learning at the start of the trial. During this period, the drivers are expected to be at their most alert, whereas, from this point onwards, mental fatigue would have likely manifested in the drivers to some extent \cite{aakerstedt2013white}. Consequently, this first five minutes is assigned the 'alert' phase during the experiment. From around the the fourth five-minute segment onward there is a consistent increase in steering angle standard deviation, indicating an impairment in the performance of all drivers, most likely caused by the driver-related task-related fatigue in question. For this reason, data from the last three five minute segments have been assigned the 'mentally fatigued' phase during the trial. Additionally, the pre-trial and post-trial CFS scores reflect substantial increases in self-reported fatigue of the participants, with the median score post-trial equal to 29, indicating moderate fatigue. However, the pre-trial scores were also fairly elevated; the pre-trial median (23) is considered to be a score indicative of mild fatigue. This is possibly a result of the breaking of normal habits (abstinence from daily caffeine) and a prolonged set-up time (up to 1.5 hours) during which the participants are mostly idle. Nevertheless, the overall trend was indicative of increased mental fatigue after the driving task. 

\begin{figure*}[!h]
    \centering
    \includegraphics{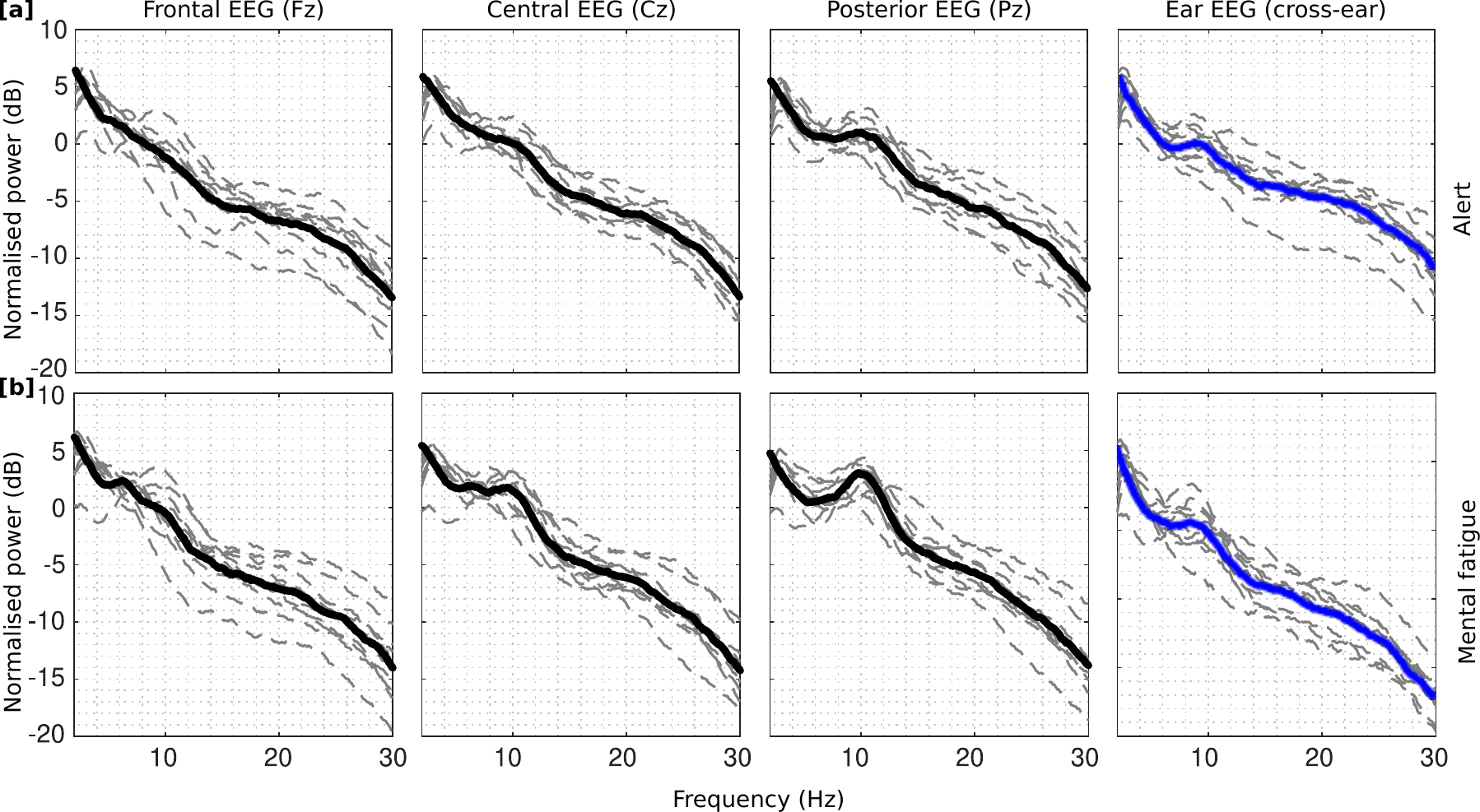}
    \caption{Mean and individual EEG spectra from ten subjects during the alert (top row) and fatigued (bottom row) phases. Individual spectra are designated in dashed gray lines, while the mean spectra from all subjects are given in a solid black (scalp EEG) or blue (ear EEG) line. The cross-ear EEG channel is compared with conventional scalp EEG channels from frontal, central, and posterior positions. Observe the characteristic increased frontal theta and posterior alpha that are commonly associated with the transition between alert and mentally fatigued phases. Such changes indicate that the experiment successfully captured a transition between alert and mentally fatigued states in the participants. The cross-ear EEG channel spectra matches the scalp EEG spectra well, while there is also evidence of increases in the theta and alpha ranges.}
    \label{fig:PSD}
\end{figure*}

Following the analysis of the behavioural and psychological signs of mental fatigue, the EEG data is evaluated. Traditionally, EEG changes in distinct frequency ranges, or bands, at specific positions on the scalp, have been analysed for the purpose of mental fatigue detection. A large amount of such empirical evidence has been accumulated in the literature, such that reliable inferences regarding mental fatigue related changes in EEG can be acquired through meta-analysis across the myriad studies. Recently, Craig \textit{et al.} conducted such a review of the literature, and concluded that the most widely reported EEG changes occur in the theta (\SI{4}{Hz} - \SI{8}{Hz}) and alpha (\SI{8}{Hz} - \SI{15}{Hz}) bands. Particularly, theta activity is dominant in the frontal regions, while alpha is most prevalent in posterior regions of the head.

Figure \ref{fig:PSD} displays the EEG spectra from the standard scalp EEG channels in the frontal (Fz), central (Cz), and posterior (POz) positions, and the proposed ear EEG (cross-ear) channel. In each plot, the spectrum from each subject is plot in gray, overlaid with the mean across all subjects, with results from the alert and mentally fatigued phases, respectively, displayed in the top and bottom row. Each individual spectra were normalised with respect to the mean power in the range \SI{1}{Hz} - \SI{30}{Hz}, in order to account for baseline increases associated with electrode impedance value. Observe the increase in the theta band in the frontal channel, and the increase in the alpha band in the posterior channel, with more moderate increases in both the theta and alpha bands in the central channel. For the proposed ear EEG, a moderate increase in both alpha and theta is evident. Overall, the spectra correspond well between the ear EEG and the standard scalp EEG.  

\begin{table}[!h]
\resizebox{\columnwidth}{!}{%
\begin{tabular}{@{}llllll@{}}
\toprule
\multirow{2}{*}{Band}           & \multirow{2}{*}{Chan.} & \multicolumn{4}{c}{$(Power_{fatigue} - Power_{alert})$}                                                  \\ \cmidrule(l){3-6} 
                                  &                          &  \textit{d}       & $95\%$ CI                        & \textit{t}        & \textit{p}            \\ \midrule
\multirow{4}{*}{Delta}  & Ear                       & 0.3  & {[}-0.3;0.7{]} & 0.93 & 0.504 \\
                                  & Fz                       & -0.1 & {[}-0.7;0.5{]} & -0.27 & 0.843   \\
                                  & Cz                       & -0.9 & {[}-1;-0.1{]}                & -2.76 & 0.059  \\
                                  &\color{blue} POz                       & \color{blue}-2                 &\color{blue} {[}-1.5;-0.7{]} & \color{blue}-6.33  & \color{blue}0.001**  \vspace{2mm}  \\ 
\multirow{4}{*}{Theta}  &\color{red} Ear                       & \color{red}1.3  &\color{red} {[}0.2;0.9{]}  & \color{red}4.03  & \color{red}0.013* \\
                                  & \color{blue} Fz                       & \color{blue} 1                  &\color{blue} {[}0.1;0.9{]}  & \color{blue}3.03   & \color{blue}0.046*  \\
                                  & Cz                       & 0.5  & {[}-0.1;0.6{]} & 1.62   & 0.224  \\
                                  & POz                       & 0                  & {[}-0.5;0.5{]} & -0.11& 0.913  \vspace{2mm} \\  
\multirow{4}{*}{Alpha} & Ear                       & 0.8  & {[}0.1;0.8{]}  & 2.6   & 0.066 \\
                                  & Fz                       & 0.2 & {[}-0.5;0.9{]} & 0.69  & 0.626   \\
                                  & \color{blue}Cz                       & \color{blue}1.3  &\color{blue} {[}0.4;1.4{]}   &\color{blue} 3.95   &\color{blue} 0.013*  \\
                                  & \color{blue} POz                       & \color{blue} 3.2  & \color{blue} {[}1;1.5{]}                   & \color{blue} 10.22   & \color{blue} <0.001***                  \vspace{2mm}  \\ 
\multirow{4}{*}{Beta} & Ear                       & -0.7& {[}-1.7;0{]}                  & -2.28 & 0.096  \\
                                  & Fz                       & -0.2 & {[}-1.7;1{]}                  & -0.55 & 0.679  \\
                                  & Cz                       & -0.4 & {[}-1.4;0.4{]}  & -1.2  & 0.382   \\
                                  & POz                      & -0.5 & {[}-1.6;0.3{]}  & -1.63  & 0.224  \\ 
\end{tabular}%
}

\caption{Results from Student's t-test for changes in EEG power between the alert and fatigued states. Effect size (Cohen's \textit{d}), \SI{95}{\%} confidence intervals, t-statistics, and p-values are provided for each EEG band of interest, and each channel (1 = cross-ear EEG, 2 = Fz, 3 = Cz, 4 = POz, ). Asterisks indicate significance for alpha $< 0.05$ (*), $< 0.01$ (**), and $< 0.001$ (***). Statistically significant results for scalp EEG are highlighted in blue, while statistically significant results fore the ear EEG are highlighted in red.}
\label{tab: stats}
\end{table}

Students t-tests were conducted in order to evaluate the statistical significance of the changes in EEG from each band in each of the EEG channels under consideration. Power values are obtained from the normalised spectra. For this analysis, the upper delta \SI{2}{Hz} - \SI{4}{Hz} and beta (\SI{15}{Hz} - \SI{30}{Hz}) bands were also analysed, since EEG studies commonly report results for these bands. Note that the delta band is usually considered to fall in the range of \SI{0}{Hz} - \SI{4}{Hz}, however, in the present study, in order to omit common low-frequency artifacts, the data was filtered with a lower cut-off frequency of \SI{1}{Hz}, rendering power values for EEG activity below \SI{2}{Hz} inaccurate. Therefore, we consider the 'upper delta' band. Effect size (Cohen's \textit{d}), \SI{95}{\%} confidence intervals (CI), t-statistics (\textit{t}), and p-values (\textit{p}), adjusted using an false discovery rate (FDR) control procedure with alpha equal to $0.05$, are displayed in Table \ref{tab: stats}. Overall, the statistical analyses reflect the widely reported changes in EEG. Increases in frontal (Fz) theta (\textit{d} = 1, CI = [0.1, 0.9], \textit{p} = 0.013) and posterior (POz) alpha (\textit{d} = 3.2, CI = [1,1.5], \textit{p} < 0.001) show significant changes between the alert and mentally fatigued phases of the trial. In addition, Cz showed an increase in alpha (\textit{d} = 1.3, CI = [0.4, 1.4], \textit{p} = 0.013), and POz showed a decrease in delta, which has been observed in rigorous EEG studies by Jap \textit{et al.} \cite{jap2009using} and Zhao \textit{et al.} \cite{zhao2012electroencephalogram}. Notably, the ear EEG demonstrates the most widely reported EEG change; a significant increase in the theta band (\textit{d} = 1.3, CI = [0.2, 0.9], \textit{p} = 0.013).  

\begin{figure*}[!h]
    \centering
    \includegraphics{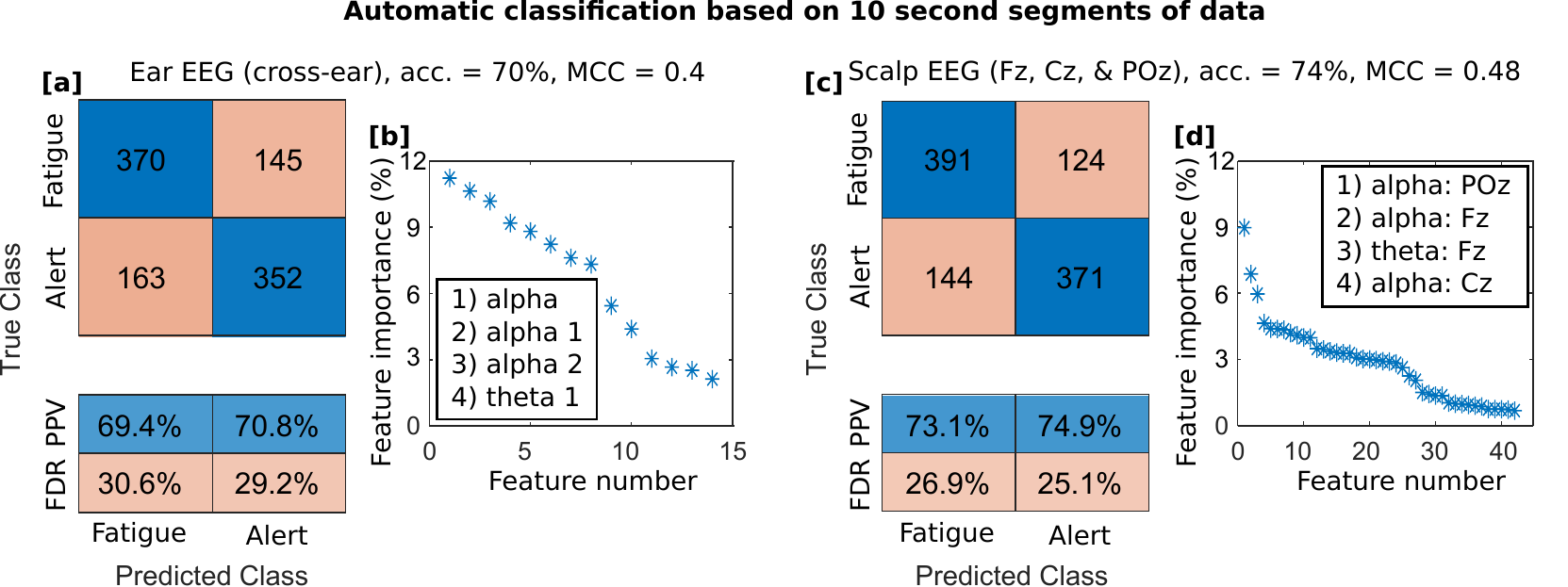}
    \caption{Confusion charts and feature importance results for the automatic EEG based driver fatigue monitoring. a) Confusion chart for the single channel, ear EEG (cross-ear) cross-validation. Each tile reflects the number of epochs which were correctly/incorrectly identified as fatigue (top left/bottom left) or alert (bottom right/top right) phases. The positive predictive value (PPV) and false discovery rate (FDR) are also displayed for each confusion chart. b) The ranked feature importance for the 14 ear EEG band features. The four most important features are displayed in the text box within the graph axes. c) Confusion chart for the multi-channel scalp-EEG model (Fz, Cz, and POz), and d) the feature importance graph for 42 multi-channel scalp EEG features.}
    \label{fig: CV}
\end{figure*}

\subsection{Automatic fatigue monitoring via ear EEG}
A more challenging, online driver fatigue detection scenario is considered next; whereby short periods of EEG recordings are evaluated for signs of fatigue, through implementation of a machine learning model based on EEG band power features. EEG data from both the alert and mentally fatigued phases were segmented in to \SI{10}{s} long epochs (\SI{50}{\%} overlap). Classification was conducted in MatLab, through the Classification Learner Application. A bagged decision tree model (Logitboost algorithm) was used for binary classification of alert and mentally fatigued EEG segments. Model hyper-parameters were optimised for the entire data set of cross-ear EEG and scalp EEG data; these were the maximum number of splits (15), number of learners (55), and learning rate (0.86). The entire dataset was used during optimisation with a view of obtaining a model trained on the largest possible dataset. Validation accuracy and Matthews Correlation Coefficient (MCC \cite{matthews1975comparison}) were used to evaluate the classification in a five-fold cross validation procedure. The MCC values reflect the true- and false-positive and negative predictions of a binary classifier, in contrast to techniques such as the F1 score. The MCC values range from -1 to 1, where -1 indicates perfect disagreement between the class labels and the data, 0 reflects no agreement, and 1 reflects perfect agreement, as with Pearson's correlation coefficient \cite{pearson1895vii}. 

Since the beta band is not strongly associated with mental fatigue and, on the other hand, it is associated with noise arising due to muscular activity (EMG), the beta band was excluded from the analysis. For the theta and alpha bands, the upper and lower half of the range were treated separately, whereas for the delta band, only the upper delta range was utilised in order to exclude filter artifacts. Two EEG features per frequency band of interest were calculated: the frequency at which the maximum power occurs, and the mean power across the entire band. In this way, for each channel, fourteen features were calculated. 

Figure \ref{fig: CV} displays the confusion charts and feature importance graphs for the cross-validation procedure. Overall, the ear EEG was able to provide good classification accuracy (acc. = 70\%) and MCC (0.4), which is comparable to a three-channel scalp EEG set-up (acc. = 74\%, MCC = 0.48). For the cross-ear EEG, multiple features contributed well to the model predictions (reflected in the shallow curve in Figure \ref{fig: CV}b)). All of the top features were mean power features, as opposed to the peak frequency features. For the scalp EEG, posterior alpha activity was the most significant feature, however, as with the cross-ear EEG model, mutliple mean power features contributed well to the model. The relatively high spread of feature importance for both the ear EEG and scalp EEG, and the high learning rate hyperparamater (0.86) indicate that a large proportion of the EEG features carry useful information. With regard to single channel performance for the scalp EEG, the cross-ear EEG performed approximately on par with the scalp channels; Fz (acc. = 69\%, MCC = 0.39), Cz (acc. = 68\%, MCC = 0.68), POz (acc. = 70\%, MCC = 0.4).

\section*{Conclusion}

Driver fatigue monitoring is an emerging technology for the detection of signs of mental fatigue in a driver. Indeed, if undetected, driver fatigue may cause lapses in vigilance, the onset of sleep, and can lead to serious road accidents. Wearable brain monitoring devices hold much promise as a result of their ability to provide a direct measure of the mental state of the driver in an efficient and seamless manner. While multiple studies have established the feasibility of scalp based wearable EEG monitoring systems, a less cumbersome and obtrusive solution is required for real-world implementation. Ear based EEG monitoring, commonly referred to as ear EEG, offers such a solution, however, the feasibility study of identifying signs of mental fatigue from a stand-alone ear EEG devices is yet to be undertaken. For the first time, through measurements on ten subjects in a simulated monotonous driving task, this study has demonstrated the ability of ear EEG to detect the most widely reported increased theta activity during the transition from alert to mentally fatigued states. This study also contributes valuable statistics for the delta, theta, alpha, and beta changes in conventional scalp EEG from key frontal, central, and posterior locations, as well as the proposed ear EEG channel. Distributions of steering wheel manoeuvre throughout the trial and subjective self assessment scores before and after the onset of driving related mental fatigue are also provided. Furthermore, ear EEG has enabled automatic real-time detection of mental fatigue based on \SI{10}{s} measurement periods, and has been demonstrated to perform well and on par with conventional scalp EEG. Importantly, only physiologically meaningful features have been considered during this machine learning procedure. Future studies should aim to increase the numerical evidence for ear EEG based mental fatigue monitoring research by investigating the characteristic scalp EEG band changes that are reflected in ear EEG from a larger and more diverse cohort of subjects. Moreover, the development of a more sensitive, machine learning based automatic detection model should also be addressed. This study has conclusively provided a new perspective on the feasibility of an ultra-wearable driver fatigue monitoring system, the so-called hearables paradigm, opening new avenues for real-world, readily implementable road-accident prevention.  

\section*{Acknowledgment}
This work was partially supported by the MURI/EPSRC grant EP/P008461 and the USSOCOM grant EESB P85655.




\bibliographystyle{IEEEtran}
\bibliography{references}

\begin{thebibliography}{10}
\providecommand{\url}[1]{#1}
\csname url@samestyle\endcsname
\providecommand{\newblock}{\relax}
\providecommand{\bibinfo}[2]{#2}
\providecommand{\BIBentrySTDinterwordspacing}{\spaceskip=0pt\relax}
\providecommand{\BIBentryALTinterwordstretchfactor}{4}
\providecommand{\BIBentryALTinterwordspacing}{\spaceskip=\fontdimen2\font plus
\BIBentryALTinterwordstretchfactor\fontdimen3\font minus
  \fontdimen4\font\relax}
\providecommand{\BIBforeignlanguage}[2]{{%
\expandafter\ifx\csname l@#1\endcsname\relax
\typeout{** WARNING: IEEEtran.bst: No hyphenation pattern has been}%
\typeout{** loaded for the language `#1'. Using the pattern for}%
\typeout{** the default language instead.}%
\else
\language=\csname l@#1\endcsname
\fi
#2}}
\providecommand{\BIBdecl}{\relax}
\BIBdecl

\bibitem{transport_tech_rep_2016}
{Department for Transport}, ``Reported {R}oad {C}asualties {G}reat {B}ritain:
  2016,'' { Tech. Rep. September}, 2017.

\bibitem{bioulac2017risk}
S.~Bioulac, J.-A.~M. Franchi, M.~Arnaud, P.~Sagaspe, N.~Moore, F.~Salvo, and
  P.~Philip, ``Risk of motor vehicle accidents related to sleepiness at the
  wheel: {A} systematic review and meta-analysis,'' \emph{Sleep}, vol.~40,
  no.~10, 2017.

\bibitem{aakerstedt2013white}
T.~{\AA}kerstedt, C.~Bassetti, F.~Cirignotta, D.~Garc{\'\i}a-Borreguero,
  M.~Gon{\c{c}}alves, J.~Horne, D.~L{\'e}ger, M.~Partinen, T.~Penzel, P.~Philip
  \emph{et~al.}, ``White paper:" sleepiness at the wheel,'' \emph{ASFA (French
  Motorways Company) and INSV (National Institute of Sleep and Vigilance)},
  p.~69, 2013.

\bibitem{saxby2008effect}
D.~J. Saxby, G.~Matthews, E.~M. Hitchcock, J.~S. Warm, G.~J. Funke, and
  T.~Gantzer, ``Effect of active and passive fatigue on performance using a
  driving simulator,'' in \emph{Proceedings of the Human Factors and Ergonomics
  Society Annual Meeting}, vol.~52, no.~21.\hskip 1em plus 0.5em minus
  0.4em\relax Sage Publications Sage CA: Los Angeles, CA, 2008, pp. 1751--1755.

\bibitem{dinges1997cumulative}
D.~F. Dinges, F.~Pack, K.~Williams, K.~A. Gillen, J.~W. Powell, G.~E. Ott,
  C.~Aptowicz, and A.~I. Pack, ``Cumulative sleepiness, mood disturbance, and
  psychomotor vigilance performance decrements during a week of sleep
  restricted to 4--5 hours per night,'' \emph{Sleep}, vol.~20, no.~4, pp.
  267--277, 1997.

\bibitem{horne1995sleep}
J.~A. Horne and L.~A. Reyner, ``Sleep related vehicle accidents,'' \emph{BMJ},
  vol. 310, no. 6979, pp. 565--567, 1995.

\bibitem{lal2001critical}
S.~K. Lal and A.~Craig, ``A critical review of the psychophysiology of driver
  fatigue,'' \emph{Biological {P}sychology}, vol.~55, no.~3, pp. 173--194,
  2001.

\bibitem{brown1997prospects}
I.~D. Brown, ``Prospects for technological countermeasures against driver
  fatigue.'' \emph{Accident Analysis and Prevention}, vol.~29, no.~4, pp.
  525--531, 1997.

\bibitem{johns2008new}
M.~W. Johns, R.~Chapman, K.~Crowley, and A.~Tucker, ``A new method for
  assessing the risks of drowsiness while driving,''
  \emph{Somnologie-Schlafforschung und Schlafmedizin}, vol.~12, no.~1, pp.
  66--74, 2008.

\bibitem{langner2010mental}
R.~Langner, M.~B. Steinborn, A.~Chatterjee, W.~Sturm, and K.~Willmes, ``Mental
  fatigue and temporal preparation in simple reaction-time performance,''
  \emph{Acta {P}sychologica}, vol. 133, no.~1, pp. 64--72, 2010.

\bibitem{zhang2016sensitivity}
H.~Zhang, C.~Wu, Z.~Huang, X.~Yan, and T.~Z. Qiu, ``Sensitivity of lane
  position and steering angle measurements to driver fatigue,''
  \emph{Transportation {R}esearch {R}ecord}, vol. 2585, no.~1, pp. 67--76,
  2016.

\bibitem{dreissig2020driver}
M.~Dreissig, M.~H. Baccour, T.~Sch{\"a}ck, and E.~Kasneci, ``Driver drowsiness
  classification based on eye blink and head movement features using the k-{NN}
  algorithm,'' in \emph{Proc. of the IEEE Symposium Series on Computational
  Intelligence (SSCI)}, 2020, pp. 889--896.

\bibitem{fan2007yawning}
X.~Fan, B.-C. Yin, and Y.-F. Sun, ``Yawning detection for monitoring driver
  fatigue,'' in \emph{Proc. of the IEEE International Conference on Machine
  Learning and Cybernetics}, vol.~2, 2007, pp. 664--668.

\bibitem{lu2022detecting}
K.~Lu, A.~S. Dahlman, J.~Karlsson, and S.~Candefjord, ``Detecting driver
  fatigue using heart rate variability: {A} systematic review,'' \emph{Accident
  Analysis \& Prevention}, vol. 178, p. 106830, 2022.

\bibitem{lal2002driver}
S.~K. Lal and A.~Craig, ``Driver fatigue:{E}lectroencephalography and
  psychological assessment,'' \emph{Psychophysiology}, vol.~39, no.~3, pp.
  313--321, 2002.

\bibitem{nguyen2017utilization}
T.~Nguyen, S.~Ahn, H.~Jang, S.~C. Jun, and J.~G. Kim, ``Utilization of a
  combined {EEG/NIRS} system to predict driver drowsiness,'' \emph{Scientific
  {R}eports}, vol.~7, no.~1, pp. 1--10, 2017.

\bibitem{lal2003development}
S.~K. Lal, A.~Craig, P.~Boord, L.~Kirkup, and H.~Nguyen, ``Development of an
  algorithm for an {EEG}-based driver fatigue countermeasure,'' \emph{Journal
  of {S}afety Research}, vol.~34, no.~3, pp. 321--328, 2003.

\bibitem{subha2010eeg}
D.~P. Subha, P.~K. Joseph, R.~Acharya~U, C.~M. Lim \emph{et~al.}, ``{EEG}
  signal analysis: {A} survey,'' \emph{Journal of {M}edical {S}ystems},
  vol.~34, no.~2, pp. 195--212, 2010.

\bibitem{monteiro2019using}
T.~G. Monteiro, C.~Skourup, and H.~Zhang, ``Using {EEG} for mental fatigue
  assessment: {A} comprehensive look into the current state of the art,''
  \emph{IEEE Transactions on Human-Machine Systems}, vol.~49, no.~6, pp.
  599--610, 2019.

\bibitem{tran2020influence}
Y.~Tran, A.~Craig, R.~Craig, R.~Chai, and H.~Nguyen, ``The influence of mental
  fatigue on brain activity: Evidence from a systematic review with
  meta-analyses,'' \emph{Psychophysiology}, vol.~57, no.~5, p. e13554, 2020.

\bibitem{caldwell2002eeg}
J.~A. Caldwell, K.~K. Hall, and B.~S. Erickson, ``{EEG} data collected from
  helicopter pilots in flight are sufficiently sensitive to detect increased
  fatigue from sleep deprivation,'' \emph{The International Journal of Aviation
  Psychology}, vol.~12, no.~1, pp. 19--32, 2002.

\bibitem{cao2014objective}
T.~Cao, F.~Wan, C.~M. Wong, J.~N. da~Cruz, and Y.~Hu, ``Objective evaluation of
  fatigue by eeg spectral analysis in steady-state visual evoked
  potential-based brain-computer interfaces,'' \emph{Biomedical {E}ngineering
  {O}nline}, vol.~13, no.~1, pp. 1--13, 2014.

\bibitem{fan2015electroencephalogram}
X.~Fan, Q.~Zhou, Z.~Liu, and F.~Xie, ``Electroencephalogram assessment of
  mental fatigue in visual search,'' \emph{Bio-medical {M}aterials and
  {E}ngineering}, vol.~26, no.~s1, pp. S1455--S1463, 2015.

\bibitem{jagannath2014assessment}
M.~Jagannath and V.~Balasubramanian, ``Assessment of early onset of driver
  fatigue using multimodal fatigue measures in a static simulator,''
  \emph{Applied {E}rgonomics}, vol.~45, no.~4, pp. 1140--1147, 2014.

\bibitem{jap2009using}
B.~T. Jap, S.~Lal, P.~Fischer, and E.~Bekiaris, ``Using {EEG} spectral
  components to assess algorithms for detecting fatigue,'' \emph{Expert Systems
  with Applications}, vol.~36, no.~2, pp. 2352--2359, 2009.

\bibitem{tanaka2012effect}
M.~Tanaka, Y.~Shigihara, A.~Ishii, M.~Funakura, E.~Kanai, and Y.~Watanabe,
  ``Effect of mental fatigue on the central nervous system: An
  electroencephalography study,'' \emph{Behavioral and {B}rain {F}unctions},
  vol.~8, no.~1, pp. 1--8, 2012.

\bibitem{trejo2015eeg}
L.~J. Trejo, K.~Kubitz, R.~Rosipal, R.~L. Kochavi, L.~D. Montgomery
  \emph{et~al.}, ``{EEG}-based estimation and classification of mental
  fatigue,'' \emph{Psychology}, vol.~6, no.~05, p. 572, 2015.

\bibitem{golz2007feature}
M.~Golz, D.~Sommer, M.~Chen, U.~Trutschel, and D.~Mandic, ``Feature fusion for
  the detection of microsleep events,'' \emph{The Journal of VLSI Signal
  Processing Systems for Signal, Image, and Video Technology}, vol.~49, no.~2,
  pp. 329--342, 2007.

\bibitem{sommer2005fusion}
D.~Sommer, M.~Chen, M.~Golz, U.~Trutschel, and D.~Mandic, ``Fusion of state
  space and frequency-domain features for improved microsleep detection,'' in
  \emph{International Conference on Artificial Neural Networks}.\hskip 1em plus
  0.5em minus 0.4em\relax Springer, 2005, pp. 753--759.

\bibitem{casson2010wearable}
A.~J. Casson, D.~C. Yates, S.~J. Smith, J.~S. Duncan, and
  E.~Rodriguez-Villegas, ``Wearable electroencephalography,'' \emph{IEEE
  {E}ngineering in {M}edicine and {B}iology {M}agazine}, vol.~29, no.~3, pp.
  44--56, 2010.

\bibitem{li2015context}
G.~Li and W.-Y. Chung, ``A context-aware {EEG} headset system for early
  detection of driver drowsiness,'' \emph{Sensors}, vol.~15, no.~8, pp.
  20\,873--20\,893, 2015.

\bibitem{lin2014wireless}
C.-T. Lin, C.-H. Chuang, C.-S. Huang, S.-F. Tsai, S.-W. Lu, Y.-H. Chen, and
  L.-W. Ko, ``Wireless and wearable {EEG} system for evaluating driver
  vigilance,'' \emph{IEEE Transactions on {B}iomedical {C}ircuits and
  {S}ystems}, vol.~8, no.~2, pp. 165--176, 2014.

\bibitem{zhu2021vehicle}
M.~Zhu, J.~Chen, H.~Li, F.~Liang, L.~Han, and Z.~Zhang, ``Vehicle driver
  drowsiness detection method using wearable {EEG} based on convolution neural
  network,'' \emph{Neural {C}omputing and {A}pplications}, vol.~33, no.~20, pp.
  13\,965--13\,980, 2021.

\bibitem{mikkelsen2015eeg}
K.~B. Mikkelsen, S.~L. Kappel, D.~P. Mandic, and P.~Kidmose, ``Eeg recorded
  from the ear: {C}haracterizing the ear-{EEG} method,'' \emph{Frontiers in
  {N}euroscience}, vol.~9, p. 438, 2015.

\bibitem{looney2012ear}
D.~Looney, P.~Kidmose, C.~Park, M.~Ungstrup, M.~L. Rank, K.~Rosenkranz, and
  D.~P. Mandic, ``The in-the-ear recording concept: {U}ser-centered and
  wearable brain monitoring,'' \emph{IEEE {P}ulse}, vol.~3, no.~6, pp. 32--42,
  2012.

\bibitem{davies2020ear}
H.~J. Davies, I.~Williams, N.~S. Peters, and D.~P. Mandic, ``In-ear {SpO2}: A
  tool for wearable, unobtrusive monitoring of core blood oxygen saturation,''
  \emph{Sensors}, vol.~20, no.~17, p. 4879, 2020.

\bibitem{davies2022wearable}
H.~J. Davies, P.~Bachtiger, I.~Williams, P.~L. Molyneaux, N.~S. Peters, and
  D.~P. Mandic, ``Wearable in-ear {PPG}: Detailed respiratory variations enable
  classification of copd,'' \emph{IEEE Transactions on Biomedical Engineering},
  vol.~69, no.~7, pp. 2390--2400, 2022.

\bibitem{goverdovsky2017hearables}
V.~Goverdovsky, W.~Von~Rosenberg, T.~Nakamura, D.~Looney, D.~J. Sharp,
  C.~Papavassiliou, M.~J. Morrell, and D.~P. Mandic, ``Hearables: Multimodal
  physiological in-ear sensing,'' \emph{Scientific {R}eports}, vol.~7, no.~1,
  pp. 1--10, 2017.

\bibitem{von2017hearables}
W.~von Rosenberg, T.~Chanwimalueang, V.~Goverdovsky, N.~S. Peters,
  C.~Papavassiliou, and D.~P. Mandic, ``Hearables: Feasibility of recording
  cardiac rhythms from head and in-ear locations,'' \emph{Royal Society {O}pen
  {S}cience}, vol.~4, no.~11, p. 171214, 2017.

\bibitem{goverdovsky2014co}
V.~Goverdovsky, D.~Looney, P.~Kidmose, C.~Papavassiliou, and D.~P. Mandic,
  ``Co-located multimodal sensing: A next generation solution for wearable
  health,'' \emph{IEEE Sensors Journal}, vol.~15, no.~1, pp. 138--145, 2014.

\bibitem{hammourArt}
G.~M. Hammour and D.~P. Mandic, ``Hearables: {M}aking sense from motion
  artefacts in ear-{EEG} for real-life human activity classification,'' in
  \emph{Proc. of the 43rd Annual International Conference of the IEEE
  Engineering in Medicine \& Biology Society (EMBC)}, pp. 6889--6893.

\bibitem{OcchipintiArt}
E.~Occhipinti, H.~J. Davies, G.~Hammour, and D.~P. Mandic, ``Hearables:
  {A}rtefact removal in ear-{EEG} for continuous 24/7 monitoring,'' in
  \emph{Proc. of the IEEE International Joint Conference on Neural Networks
  (IJCNN)}, 2022, pp. 1--6.

\bibitem{nakamura2018automatic}
T.~Nakamura, Y.~D. Alqurashi, M.~J. Morrell, and D.~P. Mandic, ``Automatic
  detection of drowsiness using in-ear eeg,'' in \emph{2018 International Joint
  Conference on Neural Networks (IJCNN)}.\hskip 1em plus 0.5em minus
  0.4em\relax IEEE, 2018, pp. 1--6.

\bibitem{stochholm2016automatic}
A.~Stochholm, K.~Mikkelsen, and P.~Kidmose, ``Automatic sleep stage
  classification using ear-{EEG},'' in \emph{Proc. of the 38th Annual
  International Conference of the IEEE Engineering in Medicine and Biology
  Society (EMBC)}, pp. 4751--4754.

\bibitem{mikkelsen2019accurate}
K.~B. Mikkelsen, Y.~R. Tabar, S.~L. Kappel, C.~B. Christensen, H.~O. Toft,
  M.~C. Hemmsen, M.~L. Rank, M.~Otto, and P.~Kidmose, ``Accurate whole-night
  sleep monitoring with dry-contact ear-{EEG},'' \emph{Scientific {R}eports},
  vol.~9, no.~1, pp. 1--12, 2019.

\bibitem{nakamura2019hearables}
T.~Nakamura, Y.~D. Alqurashi, M.~J. Morrell, and D.~P. Mandic, ``Hearables:
  {A}utomatic overnight sleep monitoring with standardized in-ear {EEG}
  sensor,'' \emph{IEEE Transactions on Biomedical Engineering}, vol.~67, no.~1,
  pp. 203--212, 2019.

\bibitem{korber2015vigilance}
M.~K{\"o}rber, A.~Cingel, M.~Zimmermann, and K.~Bengler, ``Vigilance decrement
  and passive fatigue caused by monotony in automated driving,'' \emph{Procedia
  Manufacturing}, vol.~3, pp. 2403--2409, 2015.

\bibitem{chalder1993development}
T.~Chalder, G.~Berelowitz, T.~Pawlikowska, L.~Watts, S.~Wessely, D.~Wright, and
  E.~Wallace, ``Development of a fatigue scale,'' \emph{Journal of
  {P}sychosomatic {R}esearch}, vol.~37, no.~2, pp. 147--153, 1993.

\bibitem{zhao2012electroencephalogram}
C.~Zhao, M.~Zhao, J.~Liu, and C.~Zheng, ``Electroencephalogram and
  electrocardiograph assessment of mental fatigue in a driving simulator,''
  \emph{Accident Analysis \& Prevention}, vol.~45, pp. 83--90, 2012.

\bibitem{matthews1975comparison}
B.~W. Matthews, ``Comparison of the predicted and observed secondary structure
  of {T4} {P}hage lysozyme,'' \emph{Biochimica et Biophysica Acta (BBA)-Protein
  Structure}, vol. 405, no.~2, pp. 442--451, 1975.

\bibitem{pearson1895vii}
K.~Pearson, ``{VII}. {N}ote on regression and inheritance in the case of two
  parents,'' \emph{{P}roceedings of the {R}oyal {S}ociety of {L}ondon},
  vol.~58, no. 347-352, pp. 240--242, 1895.

\end{thebibliography}
%

%




\end{document}